# Surface lattice resonance based magneto-plasmonic switch in NiFe patterned nano-structure

H. Mbarak[1,5], S. M. Hamidi[1], V. I. Belotelov[2,3,4], A. I. Chernov[2,3], E. Mohajerani[1], Y. Zaatar[5]

[1]Laser and plasma Research Institute, Shahid Beheshti University, G. C. Tehran, Iran
[2]Vernadsky Crimean Federal University, 4 Vernadskogo Prospekt, Simferopol, 295007, Russia.
[3]Russian Quantum Center, Skolkovo innovation city, Moscow, 121205, Russia
[4]Faculty of Physics, Lomonosov Moscow State University, Moscow, 119991, Russia
[5]Faculty of Sciences 2, LPA, Lebanese University, BP 90656, Jdeidet, Lebanon

Corresponding author: m_hamidi@sbu.ac.ir

**Abstract**

In this work, a 2D magneto-plasmonic grating structure combining materials with ferromagnetic and plasmonic properties is demonstrated. NiFe composite ferromagnetic material, as an active medium with tunable physical properties, and Au metal, as a plasmonic excitation layer, were the materials of choice. Here, we have experimentally investigated the active control of the plasmonic characteristics in Au/NiFe bilayer by the action of an external magnetic field, as well as the switching effect of the system. The active plasmonic control, can be achieved by the magnetization switching of the ferromagnetic material, opening a new path in the development of active plasmonic devices. To our best knowledge, this is the first demonstration of such a magneto-optical plasmonic switch based on the coupling of plasmons with magneto-optical active materials, in which the response time was estimated to be in the range of microseconds.

# Introduction:

The demand to develop high-performance active plasmonic devices for different applications has become an important need for solving many prominent problems encountered by our world. [1]. The advanced research in active plasmonics relies on finding materials that have the ability to tune plasmonic properties of metallic nanostructures through external control mechanisms [2]. Further, magneto-plasmonic systems combining periodic plasmonic arrays with magnetic materials are capable to enhance the magneto-optical responses [3,4] and to control the plasmonic properties via an external magnetic field, enabling their use in controllable plasmonic components [5,6]. Actually, different groups have been focused on the enhancement of the magneto-optical activity of the systems [7-11] as well as a few have been concentrated on the active control of the plasmon resonance [12] of such a system. In order to expand the area of research in active plasmonics and to complement our previous works that are based on electro-optical [13] and all-optical [14] effects in two dimensional (2D) plasmonic grating, an idea of improving the response of that system and controlling the plasmonic excitations as well as detecting the switching process is demonstrated by us, using the magneto-optical effects. Since our 2D plasmonic structure considered in here is a periodic array of gold nanoparticles, it supports surface lattice resonance (SLR) that results from the coupling of the diffractive waves and the localized plasmon resonance associated with individual nanoparticles at a wavelength comparable to their period.

To directly affect these plasmon modes and therefore the optical plasmonic properties of the sample, surrounding the metal nanostructures with materials that are used as tunable dielectric media was suggested. [15,16]. Consequently, modifying the refractive index of these surrounding media through the use of external fields is an important approach to achieve the control of SLR, which in turn represents the working principle of our active plasmonic device.

We demonstrate the first time the use of NiFe as an active medium for controlling the plasmonic properties. NiFe was chosen due to its low coercivity, small saturation field and high permeability [17]. By dynamically tuning its magnetization under an external magnetic field a change in the refractive index is induced, and therefore an active control of the system can be easily realized. As it was previously demonstrated our 2D plasmonic sample is characterized by its simple fabrication, feasible price, good efficiency and the possibility to be easily reproduced at high scales. Also, it was used by others in different domains of plasmonics, such as random lasing and biomedical substrate [18-19]. In this paper, by combining the plasmonic properties of 2D plasmonic structures with NiFe magnetic surrounding, we were able to obtain the active control over the SLR and thus thus the switching effect of a 2D magneto-plasmonic grating. This magneto-plasmonic structure conveys a new path to develop active plasmonic devices, which can be useful in important applications of switching and sensing.

**Experimental work:**

A two-dimensional (2D) magneto-plasmonic nanostructure was prepared by nanoimprint lithography method, see Figure 1 (a). For this purpose, the poly-dimethylsiloxane (PDMS) substrate was patterned as 2D substrate and covered by gold layer of 35 nm thickness to keep the valley part of the sample and reach to the patterned nanostructure after gold and also NiFe covered layer. The same fabrication process was used in our previous works [13,14]. The geometrical parameters of this sample are shown in Figure 1 (b) in a representative SEM image of the 2D plasmonic grating in a simple cubic structure with 1.2 µm unit cell, 2.84µm × 2.84 µm grating period, 302 nm in length and 30 nm in the diameter for gold nanowire. A lateral cross section of the PDMS-based sample is presented in Figure 1 (c), which clearly shows the two-dimensional grating of the sample.

After the deposition of the gold layer on the PDMS substrate, a magnetic NiFe thin film is sputtered on. The sputtering is realized using magnetron sputtering deposition with a thickness of 30 nm. Afterwards, the longitudinal Kerr effect geometry was considered by applying a 30 mT DC magnetic field in the plane of the sample and parallel to the incidence plane.

Thereafter, the reflectance spectra of the sample were recorded for positive and negative directions of magnetic field in rotational analyzer method, for p polarization light and at a specific incidence angle of 54 degrees. Then, the switching manner of the sample was measured under external magnetic field in "ON" and "OFF" states using spectral longitudinal magneto-optical Kerr effect measurement setup, and then we have recorded the switching manner behavior under ON and OFF states for both directions (right and left) of the magnetic field, separately (Fig. 2).

Optical characterization is done through spectroscopy measurement by using an unpolarized light of a Xe lamp ($\lambda = 400–800$ nm). Reflected light from the sample is coupled into an optical fiber and then recorded using UV-visible spectrometer (HR4000G-UV-NIR from Ocean Optics) [13]. All experiments are performed at room temperature.

Finally, the spectral magneto-optical Kerr rotation in the longitudinal configuration was recorded by our home-made MO setup [23] which covers full visible region of light.

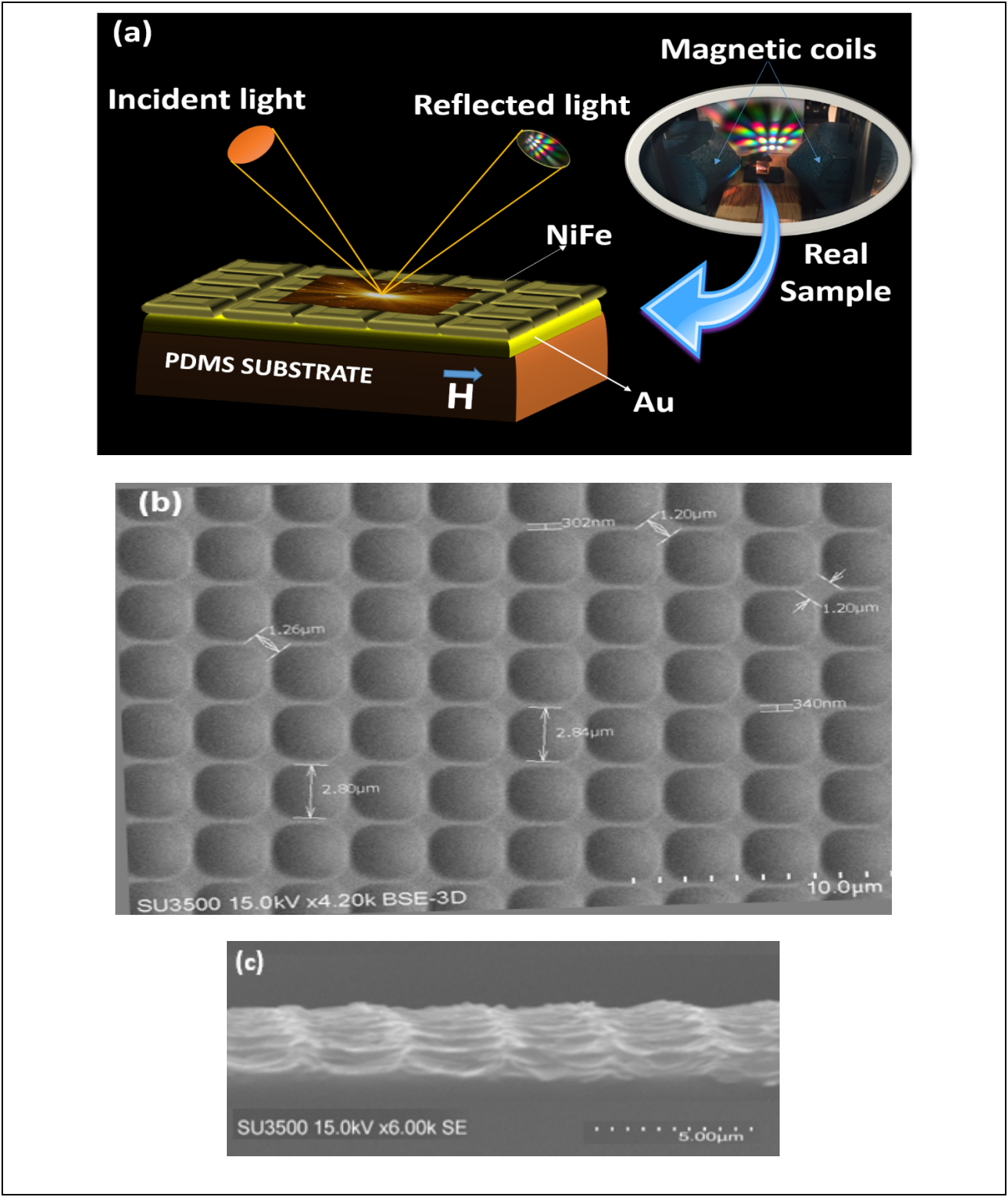

*Figure 1. (a) Sketch of the 2D magneto-plasmonic grating in the L-MOKE configuration. (b) SEM image of the sample (c) cross section of the sample placed on PDMS substrate*

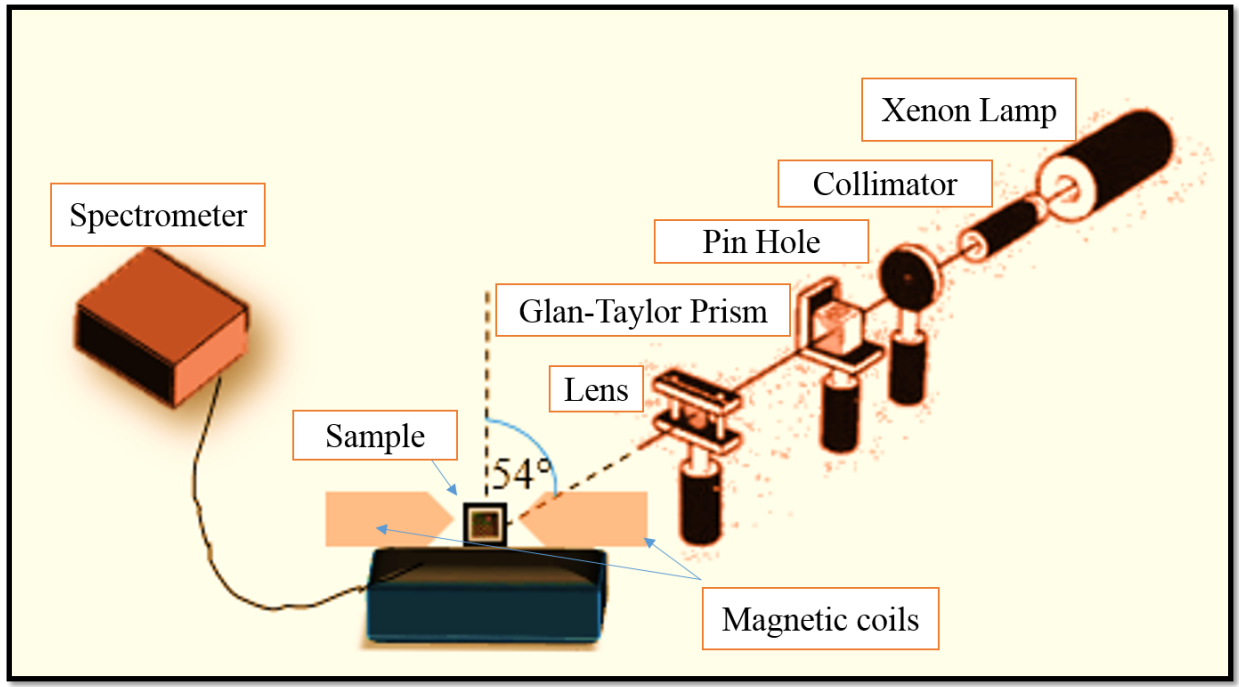

*Figure 2. Schematic diagram of the experimental setup for switching response.*

## III. Results and Discussion:

Figure 3 shows the magnetic hysteresis loops of the Au/NiFe film deposited on a PDMS substrate and measured using the vibrating-sample magnetometer (VSM) It indicates that the created 2D sample is ferromagnetic and it can be saturated at 1 Tesla for out of plane field.

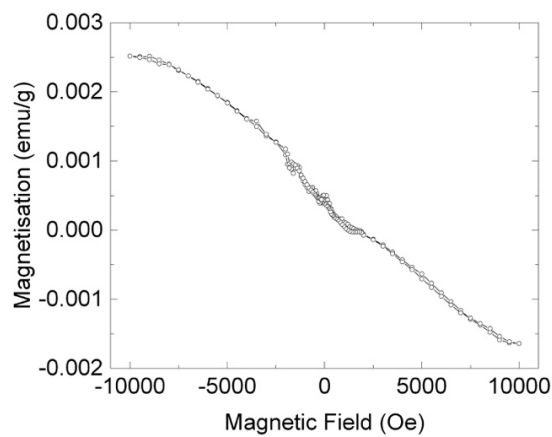

*Figure 3: the VSM measurement of the fabricated 2D sample.*

Since our sample is a periodic array of metallic nanoparticles that are physically separated by a period comparable to the wavelength of the incident light, a remarkably narrow plasmon resonances also referred as surface lattice resonance (SLR) can be generated as a collective response resulting from the coupling of diffracted waves and localized surface plasmon resonances (LSPR) of individual nanorods in far-field measurements. The LSPRs, associated with individual nanorods, are very sensitive to the variation of the dielectric function of the surrounding medium and therefore any variation of the refractive index can affect the coupling between LSPRs and a diffraction orders (DOs) causing a change in the SLRs.

This important relationship between SLR and the affecting factors can be considered as an essential basis to achieve the active plasmon control which results in modulation depth and shift of the reflected light in the sample.

Based on our knowledge, the SLR-based 2D grating should be excited at a single incidence angle to satisfy the condition coupling between the diffracted waves and LSPRs. In our structure, the excitation of SLR was proved at 54° and the reflectivity spectrum was investigated at (-1, 0) diffraction order related SLR.

As shown in Figure 4(a), SLRs are recorded in three different sates: without magnetic field and with magnetic field for right and left field directions. As it can be noticed, two peaks associated to SLRs of the array occur at $\sim 608\ nm$ and $\sim 697\ nm$ as well as the diffraction order wavelength is observed at $\lambda_{DO} = 650\ nm$. When the magnetic field is applied, a blueshift of about $\sim 6nm$ can be observed, confirming the sensitive dependence of SLR on its surrounding environment. Apart from that, amplitude of the both SLR peaks changes with applied magnetic field. To be sure that the obtained shift is caused by NiFe under the external magnetic field, reflectance spectra of the original sample (2D gold grating without NiFe) are recorded without and with magnetic field (see

Figure 4 (b)). This figure shows that the spectral position of the resonance wavelength didn't shift in the presence of the magnetic field, proving that the active control of the plasmonic structure (blue shift) is achieved owing to the surrounding NiFe medium.

Indeed, when an external magnetic field is applied the magnetization of the ferromagnetic material will be induced, causing a change in the refractive index. This change of the effective refractive index represents the essential way to tune the surface lattice resonance, which is calculated by derivation of the sample effective refractive indices with and without magnetic field by Kramers-Kronig relation [20] as shown in the inset of Figure 4 (a). We should note that this effect is even in magnetization and therefore, we observe no difference for right and left directions of the magnetization.

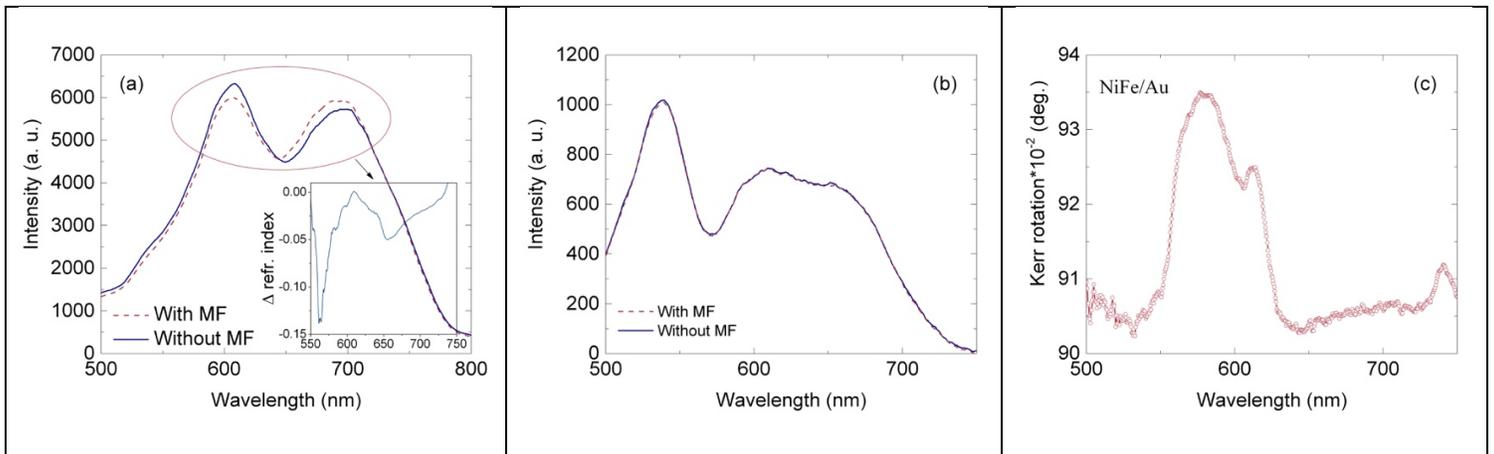

*Figure 4: (a) Reflectance spectra of the sample with NiFe layer with (dash line) and without magnetic field (solid line); the inset shows the difference in the real part of the effective refractive index with and without magnetic field in the SLR region (b) Reflectance spectra of the sample without NiFe layer without and with magnetic field and (c) magneto-optical Kerr rotation spectrum in the longitudinal configuration.*

According to Mie theory, any variation of the dielectric constant or refractive index of the surrounding medium will affect the LSPR and therefore the coupling between them (LSPR) and the diffracted waves will be affected, causing a change in SLRs. The sensitive dependence of SLR on its surrounding environment forms an essential basis for active control of SLR. As a result, if the magnetic field is applied, the refractive index decreases and therefore owing to Mie theory, if we have reduction of the refractive index, we should obtain a blueshift in the spectrum, which is clearly shown in the experimental data.

In addition, the spectral MO Kerr effect in the visible region confirms enhanced MO rotation in the vicinity of SLR wavelengths as shown in Figure 4 (c). It must be mentioned that in other two configurations as transverse and also polar one, we don't have MO response of the sample.

On the other hand, the magnetization switching was done by applying a static magnetic field to the sample [21, 22]. To get the switching response, we applied the magnetic field larger than the coercive field because it should have enough anisotropy for magnetization switching. For this purpose, as we mentioned in the experimental part, a spectral magneto-optical Kerr setup in longitudinal configuration was used.

After clarification of suitable plasmonic response, we focused on the resonance dip around ~ 649 nm and then recorded the switching behavior under "ON" and "OFF" states of the magnetic field. Actually, in the "ON" state of magnetic field the reflectance intensity is increased, whereas when the magnetic field is turned off a decrease of the intensity can be observed. This intensity variation induced by the magnetic field switching is mutually related to the blueshift observed above in Figure 4 (a), giving rise to the important dependence of NiFe on the plasmonic structure properties.

The reversible switching process under magnetic field in the longitudinal configuration clarified the reversibility and the reliability of the refractive index change of the magneto-plasmonic sample (Figure 5(a)). The response time of the switching process is considered to be 40 µs for surface lattice resonance of the sample as shown in Figure 5 (b). This main measurement has been carried out by fixing the magnetic field direction and recording the optical response at the resonance wavelength at on and off state of magnetic field.

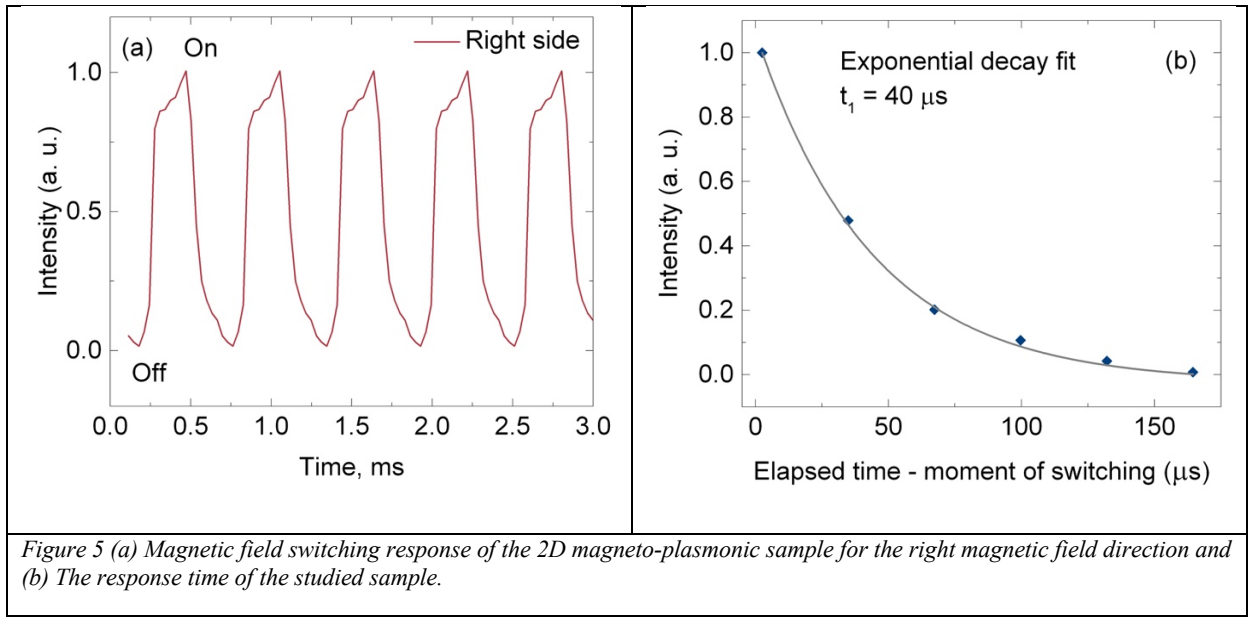

*Figure 5 (a) Magnetic field switching response of the 2D magneto-plasmonic sample for the right magnetic field direction and (b) The response time of the studied sample.*

This fast switching response in the vicinity of coercive field was confirmed by the soft ferromagnetic sample which has a very narrow loop as shown in VSM result. It means that the sample was fixed in the easy axis magnetization direction and excited in the easy axis direction.

In our experiments, due to the importance of exciting the SLR, the incident angle of light was 54°, and because the longitudinal MOKE signal is proportional to the sine of this angle, the effect of the in-plane component of magnetization is even larger which can be affected onto the switching time of the magnetization of the sample and thus onto the effective refractive index and SLR. This MOKE is shown in Figure 4 (d) which confirms the enhanced rotation at the SLR wavelength and at this resonance incidence angle.

## IV. Conclusion

In summary, a switchable two-dimensional magneto-plasmonic grating based on patterned NiFe nanostructure is reported. The grating design was inspired by merging the plasmonic properties of the gold array and the optical properties of the composite NiFe. The periodic array of gold nanostructure is fabricated on PDMS substrate using nanoimprint lithography method and then covered by a thin layer of NiFe using magnetron sputtering deposition. We find that the optical response of the 2D lattice of Au nanoparticles shows a narrow plasmon resonances, also called "surface lattice resonance (SLR)". Results indicate that the spectral position of the resonance wavelength of the SLRs can be dynamically tuned by changing the magneto-optical properties of the surrounding ferromagnetic material (NiFe) under an external magnetic field. Also, the magnetic field-induced 2D plasmonic grating switching time was measured and estimated to be in the order of microseconds. This new type of active plasmonic heterostructure can be used for future optoelectronic devices such as magneto-plasmonic switches by controlling the plasmonic properties due to variation of the dielectric tensor elements via external magnetic field.

## Acknowledgements

VIB and AIC acknowledge the financial support by Russian Foundation for Basic Research (grant # 18-29-02120 and 19-02-00856).